\makeatletter
\@ifundefined{@parse@version@dash}{%
\def\@parse@version#1{\@parse@version@0#1}
\def\@parse@version@#1/#2/#3#4#5\@nil{%
\@parse@version@dash#1-#2-#3#4\@nil}
\def\@parse@version@dash#1-#2-#3#4#5\@nil{%
  \if\relax#2\relax\else#1\fi#2#3#4 }
}{}
\makeatother

\documentclass[%
 reprint,
 amsmath,amssymb,
 aps,
]{revtex4-2}

\usepackage{multirow}
\usepackage{graphicx}
\usepackage{epsfig}
\usepackage{dcolumn}
\usepackage{bm}
\usepackage{hyperref}
\usepackage[mathlines]{lineno}
\usepackage{xcolor}

\begin{document}

\preprint{APS/123-QED}

\title{The generation and regulation of public opinion on multiplex social networks}

\author{Zhong Zhang}
 \altaffiliation{School of Mathematics, Shandong University, Jinan 250100, Shandong, China.}
\author{Jian-liang Wu}
\altaffiliation{School of Mathematics, Shandong University, Jinan 250100, Shandong, China.}
\author{Cun-quan Qu}
\altaffiliation{Data Science Institute, Shandong University, Jinan, 250100, P.R.China.}
\author{Fei Jing}
\altaffiliation{School of Data Science, City University of Hong Kong, Hong Kong SAR, China.}
\email{feijing@cityu.edu.hk}


\begin{abstract}
The dissemination of information and the development of public opinion are essential elements of most social media platforms and are often described as distinct, man-made occurrences. However, what is often disregarded is the interdependence between these two phenomena. Information dissemination serves as the foundation for the formation of public opinion, while public opinion, in turn, drives the spread of information. In our study, we model the co-evolutionary relationship between information and public opinion on heterogeneous multiplex networks. This model takes into account a minority of individuals with steadfast opinions and a majority of individuals with fluctuating views. Our findings reveal the equilibrium state of public opinion in this model and a linear relationship between mainstream public opinion and extreme individuals. Additionally, we propose a strategy for regulating public opinion by adjusting the positions of extreme groups, which could serve as a basis for implementing health policies influenced by public opinion.
\end{abstract}

\maketitle


\section{\label{sec:level1}INTRODUCTION}

With the advancements in modern media tools and the proliferation of social media, the exchange of information, views, and emotional expressions among people has become more frequent and varied. The simultaneity of information spreading and opinion evolution plays a significant role in shaping each other. Public opinion is subject to frequent changes with the influx of new information, while groups with differing opinions demonstrate varying levels of enthusiasm for disseminating information. These aspects, which are related to emotional expression, public opinion formation, and the dynamic relationship between information dissemination and public opinion, give rise to the development of a compound dynamics model that incorporates networks based on human interactions \cite{re1,re2,re3,re4,re5,re6}. These models allow for the simultaneous characterization of various social diffusion behaviors, such as the spread of infectious diseases \cite{re7,re8,re9,re10,re11}, human collective behavior \cite{re12,re13,re14,re15}, the popularity of opinions \cite{re16,re17,re18}, and the generation of hate speech and extremism \cite{re19}, in both social and biological sciences \cite{re20,re21}.

In many studies on the diffusion of social behavior, information spreading is considered a special case of social contagion and is often modeled using epidemics dynamics, such as the susceptible-infectious (SI), susceptible-infectious-recovered (SIR), and susceptible-infectious-susceptible (SIS) models~\cite{re22}.
Recently, researchers have shifted their focus towards combining empirical data with epidemic dynamics. Several studies combine empirical data with the SI model to investigate the evolution of propagation in networks \cite{re23,re24}.
Recent studies on information dissemination tend to be more empirical~\cite{re25,re26} and place a greater emphasis on real-time information dissemination pathways~\cite{re27} and the trend of rumor spreading~\cite{re28,re29,re30}.

Simultaneously, the real-time evolution of public opinion has caught the attention of both physicists and sociologists.
Given the prevalence of extreme speech, fanaticism, and populism on social media~\cite{re31}, the complex and dynamic evolution of public opinion holds significant weight in political elections, product promotions, and other fields~\cite{re32,re33,re34}.
Early researches on public opinion primarily explored three aspects: individual level, model mechanism, and network structure. At the individual level, researchers examine complex and diverse individual characteristics such as personal preferences~\cite{re35}, substitution~\cite{re36}, personalized information~\cite{re37}, and extreme individuals~\cite{re38, re39} to gain insight into the generation of public opinion.
Furthermore, researches based on model mechanisms provide insight into the evolution laws of public opinion. 
For instance, several studies consider the critical behavior of nodes under varying contact conditions\cite{re40}, investigate the impact of media quality\cite{re41} and explore public opinion polarization and the evolution of global consensus across various topics \cite{re42}.
Meanwhile, researches based on network structure shed light on the impact of network topology on the evolution of public opinion~\cite{re43, re44}

Recently, more and more researchers realized that information spreading and public opinion formation are two simultaneous and interacting processes that should be integrated into a composite model.
Therefore, some researchers propose a multi-layer network dynamics framework that considers both the network structure and model mechanism \cite{re45}. In addition, some studies explore the integration of individual-level cognitive biases and platform interactions to analyze public opinion and the of echo chambers \cite{re46}.
Although the establishment of a composite model can provide us with a deeper insight into social behavior cognition, the composite model itself is more complex, difficult to analyze and simulate, and difficult to verify on real data. These difficulties and importance make it necessary for us to further explore the analytical techniques and mechanism laws of the composite model.

In this paper,
we propose an information-opinion model on a duplex network that incorporates both the spread of information and the dynamics of opinion.
We develop an analytical technique based on the heat conduction model that successfully resolved the steady-state consistent distribution of this composite model. Through simulation experiments, we find the intrinsic connection between public opinion distribution and network structure. We then use quenched mean-field techniques to build a mathematical representation of public opinion, the mean of the steady-state public opinion distribution. Finally, on the basis of perturbation analysis, the relationship between node influence and public opinion is built, and a corresponding public opinion regulation strategy is proposed.

\section{Information-opinion model on duplex network}
\begin{figure}
\centering
\includegraphics[width=\linewidth]{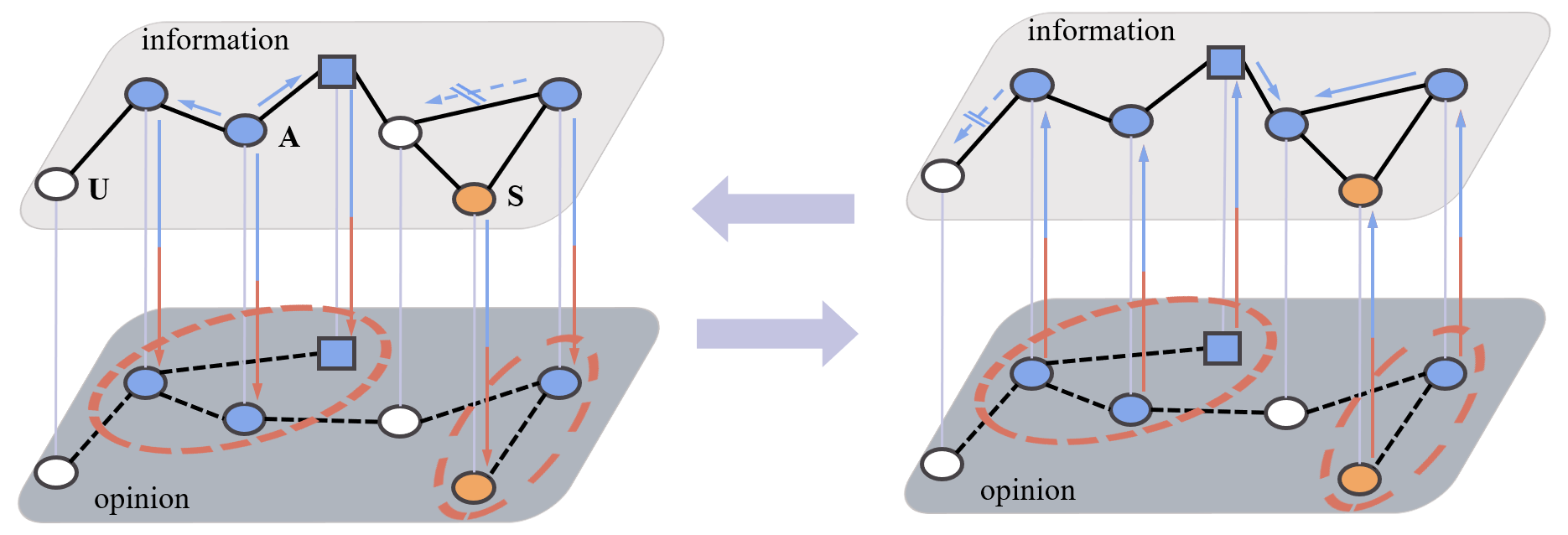}
\caption{Schematic illustration of information-opinion dynamics in population with information layer and opinion layer.
In the process of information spreading, all individuals can be divided into unknown individuals (white), active individuals (blue) and silent individuals (orange).
Additionally, the whole population can also be divided into general individuals (circle) and extreme individuals (square).
At each moment, active individuals pass the information to their unknown neighbors with a temporal probability dependent on their own opinion in the information layer, which determines the scope of opinion interaction (red circle) in the opinion layer. Then, individuals update their opinions and determine the spread probability at the next time.}
\label{fig1}
\end{figure}

In our hypothesis, information dissemination and public opinion interaction are based on the same group, but the channels for information dissemination and public opinion evolution are different.
We use a duplex network $G(V,E^I,E^O)$ to describe the network structure behind this dynamic, where $V$ includes all individuals, and $E^I$ and $E^O$ contain all links in the information layer and opinion layer, respectively (see Fig.~\ref{fig1}). Moreover, we define $A^I= \{a^I_{ij}\}$ and $A^O = \{a^O_{ij}\}$ as the adjacency matrix of the information layer and public opinion layer respectively.


In the information layer, all individuals can be divided into three types, which are the unknown ($\mathcal{U}$) state that does not know the information, the active ($\mathcal{A}$) state that knows the information and is willing to spread it, and the silent ($\mathcal{S}$) state that knows the information but does not spread it.
At time $t$, each individual $i$ with $\mathcal{U}$ state will know the information from the neighbors with $\mathcal{A}$ state (if possible) and be transformed into $\mathcal{A}$ state with probability $\beta_i^t$; each individual with $\mathcal{A}$ state will be transformed into $\mathcal{S}$ state after propagating $\omega$ rounds. We define $\alpha_i^t \in \{\mathcal{U},\mathcal{A},\mathcal{S}\}$ to represent the information state of individual $i$ at time $t$.

In the opinion layer, individuals with $\mathcal{A}$ state or $\mathcal{S}$ state will generate their own initial opinions and interact with each other. The opinion state of individual $i$ at time $t$ is defined as $\theta_i^t \in [-1,1]$,
encompassing all possible states ranging from extreme opposition to extreme support, where the magnitude of the value represents the degree of opposition or support. 
For example, $\theta_i^t = 1$ signifies that $i$ is an extreme supporter, and $\theta_i^t = -1$ implies that $i$ is an extreme opponent. The population can be divided into general and extreme groups, represented by $N_G$ and $N_E$, respectively. The extreme group can be further divided into extreme supporter group ($N_E^+$) and extreme opponent group ($N_E^-$). All general individuals update their opinions according to the following rules:
\begin{eqnarray}\label{eq1}
	\dot{\theta_i} = \frac{1}{a^O_i}\sum\limits_{j: \alpha_j^t \neq \mathcal{U}} {a^O_{ij} (\theta_j - \theta_i)},
\end{eqnarray}
where~$a^O_i$~is the number of neighbors of $i$ participating in the opinion interaction.

The Fermi distribution function and prospect theory suggest that individuals are more likely to take action for behaviors they identify with. In addition, the relationship between information dissemination probability and opinion interaction demonstrated in \cite{re47, re48} follows a form similar to the Fermi distribution function. Thus, our model effectively captures the feedback mechanism that adjusts opinions during the dissemination of information, in a manner that resembles the Fermi function:
\begin{eqnarray} \label{eq2}
	\beta_i^t = \frac{2\gamma}{1 + \exp{( - h |\theta_i^{t-1}| })},
\end{eqnarray}
in which the parameters $\gamma$ and $h$ are the adjustable feedback intensity.


\section{Steady-state distribution of opinions}
Different from a single information dissemination dynamics or public opinion evolution dynamics, the information-opinion model on duplex network has a more complex behavior pattern. On the one hand, with the spread of information, more and more people know the news, which means that more and more individuals have joined the process of public opinion interaction. On the other hand, the fluctuation of individual opinions over time will feed back into the information diffusion behavior and make it more uncertain.

To address this issue, we propose a general framework for the dynamics of multi-layer networks \cite{re49} and utilize the heat conduction model and quenched mean-field techniques to explore the evolution mechanism.
In the process of information diffusion, the changes from $\mathcal{U}$ state to $\mathcal{A}$ state and from $\mathcal{A}$ state to $\mathcal{S}$ state are irreversible, that is, the evolution of individual information state can only be one-way. On the contrary, in the process of the evolution of public opinion, the evolution of personal opinions is always two-way, and any individual may swing back and forth between support and opposition. Therefore, we might as well assume that the behavior of information dissemination terminates within a finite time, while the dynamics of public opinion can evolve infinitely over time.


In the process of information dissemination, the key issue is to estimate its dissemination range, that is, the probability that each individual knows the information under the mean field.
We define $P_i^t$ as the probability that individual $i$ knows the information at time $t$, which satisfies
\begin{eqnarray}\label{eq3}
P_i^t = 1 - \prod_{j\in \Gamma^I_i}  \left (1 - P_j^{t-1}R_{j} \right),
\end{eqnarray}
where $\Gamma^I_i$ contains all neighbors of $i$ in the information layer. We also define $R_{j}$ as the probability that individual $j$ delivers the information to his neighbors within $\omega$ times, which can be approximated by the following formula,
\begin{eqnarray}\label{eq4}
R_{j} = \sum_{k=0}^{\omega} \prod_{l=0}^{k-1} \Big[1 - \beta_j^l \Big] \beta_j^k \approx \sum_{k=0}^\omega \beta_j^k.
\end{eqnarray}

Under the thermodynamic limit and single-source propagation (i.e., if $i$ is source node then $P_i^0 = 1$ otherwise $P_i^0 = 0$), we assume that $p_i = P_i^{\infty}$ is the probability that $i$ knows the news after the end of information dissemination. Therefore, let $B=\left( b_{ij} \right)_{N\times N}$ describe the possible range of information dissemination, where $b_{ij} = p_ip_ja^O_{ij}$ is the probability that both~$i$~and~$j$~are acquainted with the message and exchange opinion with each other. Therefore, Eq.~(\ref{eq1}) satisfies,
\begin{eqnarray} \label{eq5}
\dot{\theta_i} = \frac{1}{\sum_{j=1}^N b_{ij}} \sum_{j=1}^N {b_{ij} (\theta_j - \theta_i)}.
\end{eqnarray}
For all individuals, considering that extreme individuals keep their opinions unchanged, when the opinion interaction process ends, Eq.~(\ref{eq5}) can be rewritten as
\begin{eqnarray} \label{eq6}
\theta_i = \frac{1}{\sum_{j=1}^N b_{ij}} \sum_{j=1}^N b_{ij} \theta_j + f_i,
\end{eqnarray}
where $f_i$ is the external force maintaining extreme opinions, if $i\in N_G$, $f_i = 0$. Further, we can analyze public opinion as follows,
\begin{eqnarray} \label{eq7}
(I - D^{-1}B)\mathbf{\theta} = F,
\end{eqnarray}
where~$\mathbf{\theta}=\left( \theta_1,\theta_2,\cdots,\theta_N \right)^{\mathrm{T}}$~captures the public opinion of the population of $N$ individuals. $D = \left( d_{ij} \right)_{N\times N}$ in which $d_{ij}= \sum_{k=1}^N b_{ik}$ if $i = j$, otherwise $d_{ij} = 0$. $I$ is identity matrix and $F=\left(f_1,f_2,\cdots,f_N\right)^{\mathrm{T}}$.
Note that, as demonstrated in \cite{re49}, Eq.~(\ref{eq7}) is the discrete analog of Fourier's Law and conforms to the paradigm of heat conduction.
Since $D^{-1}B$ is a Markov matrix, we know that~$I-D^{-1}B$~is irreversible. Similar to \cite{re49}, we define that~$\mathbf{u}$~and~$\mathbf{v}$~are the right eigenvector and left eigenvector of~$D^{-1}B$, corresponding to eigenvalue~$1$, respectively.
Then, $I - (D^{-1}B - \mathbf{uv})$ is invertible. Assume $\mathbf{\theta_G}$ and $\mathbf{\theta_E}$ are the stationary opinions of general and extreme groups respectively, and $\Omega^\ast = (I - B + \mathbf{uv})^{-1}$. We split~$\mathbf{u}$, $\mathbf{v}$, $F$~and~$\Omega^\ast$~based on the division of extreme individuals and general individuals.
Let~$c = \mathbf{v_E}\mathbf{\theta_E} + \mathbf{v_G}\mathbf{\theta_G}$~, Eq.~(\ref{eq7}) can be rewritten as follows:
\begin{eqnarray}\label{eq8}
	\displaystyle \left [\begin{array}{c} \mathbf{\theta_E} \\ \mathbf{\theta_G} \end{array}\right] = \left [\begin{array}{cc} \Omega_{EE}^\ast & \Omega_{EG}^\ast \\ \Omega_{GE}^\ast & \Omega_{GG}^\ast \end{array}\right ]
	\displaystyle \left [\begin{array}{c} F_E \\ \mathbf{0} \end{array}\right] + \left [\begin{array}{c} c\mathbf{u_E} \\ c\mathbf{u_G} \end{array}\right].
\end{eqnarray}
Consequently, we can solve $\mathbf{\theta_G}$ as
\begin{eqnarray}\label{eq9}
\mathbf{\theta_G} = c\mathbf{u_G} + \Omega_{GE}^\ast (\Omega^\ast_{EE})^{-1}\mathbf{\theta_E} - c\Omega_{GE}^\ast (\Omega^\ast_{EE})^{-1}\mathbf{u_E},
\end{eqnarray}
where
\begin{eqnarray}\label{eq10}
c = \frac{\mathbf{v_E}\mathbf{\theta_E} + \mathbf{v_G}\Omega_{GE}^\ast (\Omega^\ast_{EE})^{-1}\mathbf{\theta_E}}{1 - [\mathbf{v_Gu_G} - \mathbf{v_G}\Omega_{GE}^\ast (\Omega^\ast_{EE})^{-1}\mathbf{u_E}]}.
\end{eqnarray}

As shown in Eq.(~\ref{eq9}), we analyze the evolutionary process of information dissemination and public opinion formation and derive the stationary distribution of public opinion based on two key factors: the topology of the structured population ($B$) and the distribution of extreme opinions ($\theta_E$).
The former takes into account the various channels of information dissemination and public opinion interaction within the structured population, while the latter characterizes the number and location of extreme groups and their level of extreme support or opposition.

\begin{figure}[h]
\centering
\includegraphics[width=\linewidth]{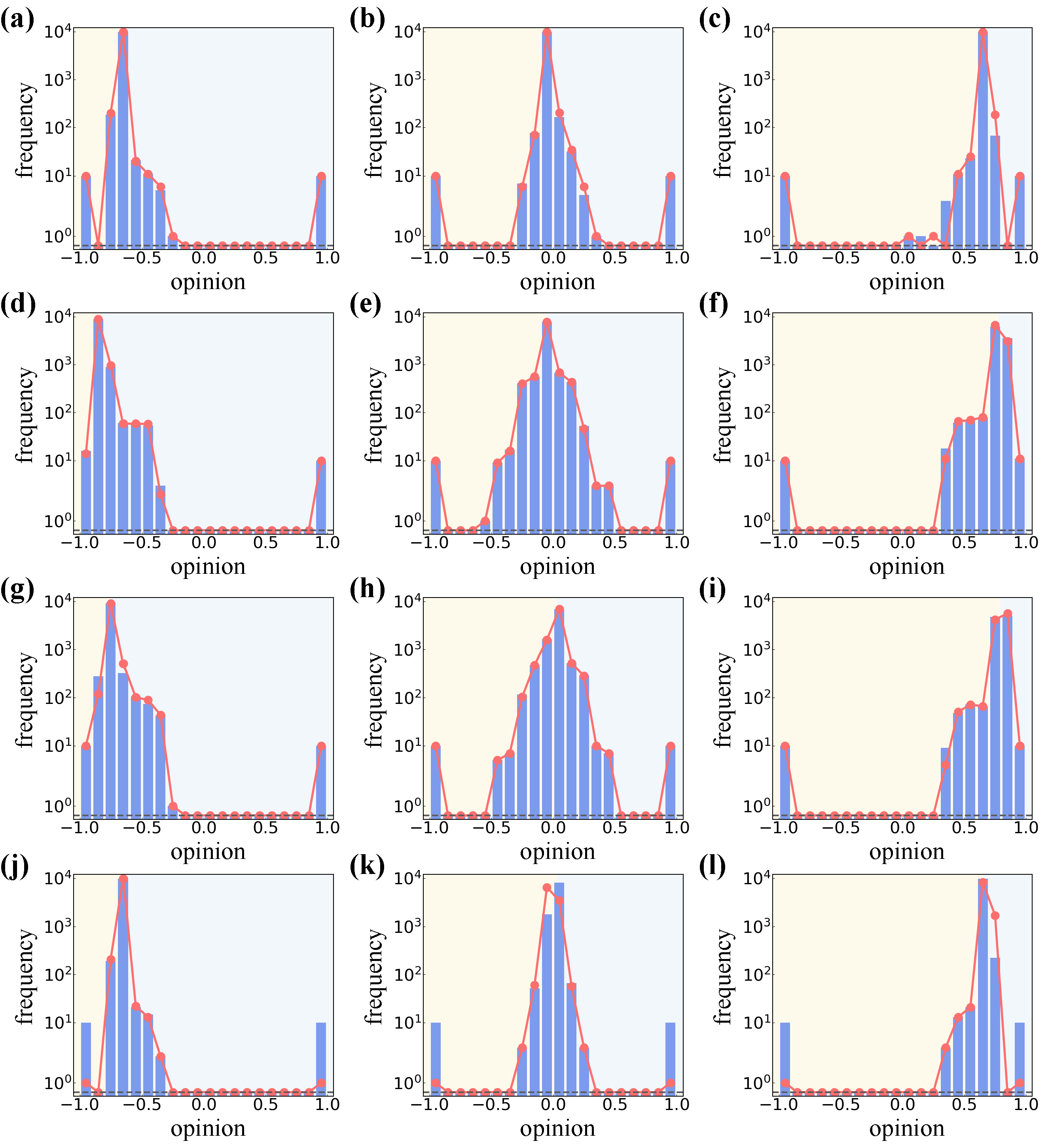}
\caption{Public opinion distribution on four synthetic networks, including BA-ER (a-c), BA-BA (d-f), ER-BA (g-i) and ER-ER (j-l).
The line and histogram represent the analytical results and simulation results, respectively.
In the experiments, we let $\omega=10$, $\gamma=0.05$, $h=1$, $C_s + F_s = 10$, and divide the graph based on the mainstream opinion. Then, tunable parameter $C_s$ belongs to $\{0,1,2,\cdots,10\}$.
However, this figure only shows the results based on three cases: $C_s=1$ (a, d, g, j), $C_s=5$ (b, e, h, l) and $C_s=9$ (c, f, i, l).
Other results are shown in Appendix~\ref{apx:2}.}\label{fig:2}
\end{figure}

\begin{figure}
\centering
\includegraphics[width=\linewidth]{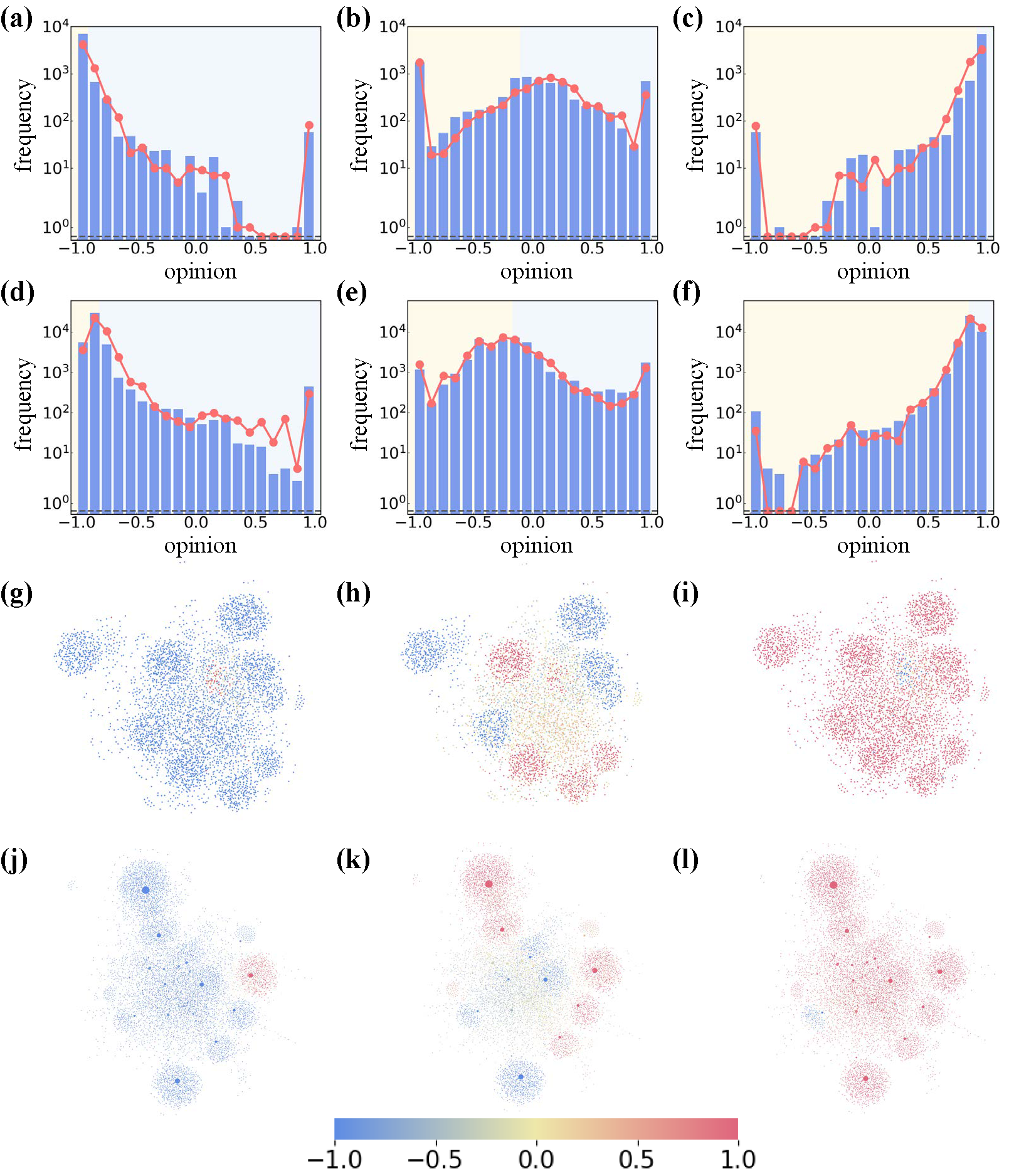}
\caption{Frequency statistics and visualization of public opinion on empirical networks, including MOS (a-c, g-i) and CAN (d-f, j-l).
The line and histogram represent the analytical results and simulation results, respectively.
In the experiments, we let $\omega=10$, $\gamma=0.05$, $h=1$, $C_s + F_s = 10$, and divide the graph based on the mainstream opinion. Then, tunable parameter $C_s$ belongs to $\{0,1,2,\cdots,10\}$.
However, this figure only shows the results based on three cases: $C_s=1$ (a, d, g, j), $C_s=5$ (b, e, h, k) and $C_s=9$ (c, f, i, l).
Other results are shown in~\ref{apx:2}.}\label{fig:3}
\end{figure}

Next, we investigate the effect of diverse multiplex network structures and extreme groups on opinion distributions. Specifically, we consider a straightforward scenario where all extreme individuals are located either at hub nodes or frontier nodes,
The extreme individuals located at hub nodes are called \emph{central extreme supporters} or \emph{central extreme opponents}, with the number of these individuals represented by $C_s$ and $C_o$, respectively. Similarly, the extreme individuals located at frontier nodes are called \emph{frontier extreme supporters} or \emph{frontier extreme opponents}, with the number of these individuals represented by $F_s$ and $F_o$, respectively.
In addition, we assume that $C_s+F_s=C_o+F_o$ and $C_s+C_o=F_s+F_o$, which allows us to examine the impact of different extreme group distributions on public opinion distribution by tuning the parameter $C_s$. Specifically, we consider three possible cases of extreme groups: extreme opponents dominant (e.g., Fig.~\ref{fig:2}(a,d,g,j) and Fig.~\ref{fig:3}(a,d)), extreme opponents and supporters are evenly matched (e.g., Fig.~\ref{fig:2}(b,e,h,k) and Fig.~\ref{fig:3}(b,e)), and extreme supporters are dominant (e.g., Fig.~\ref{fig:2}(c,f,i,l) and Fig.~\ref{fig:3}(c,f)).

The stationary opinion distributions of groups in the four different synthetic network models, are shown to have a bell-shaped distribution centered on the mainstream opinion, as depicted in Fig.~\ref{fig:2}. The opinion distribution not only reaches its maximum value at the mainstream opinion, but also presents a monotonic decrease on either side of the mainstream opinion(i.e., light yellow area and light blue area), excluding the extreme opinion which remains unchanged. Additionally, the analytical results represented by the red dot and line are in perfect agreement with the simulation results depicted by the blue histogram, indicating that our analysis correctly captures the stationary distribution of the information-public opinion dynamics.

Fig.~\ref{fig:3} presents the stationary opinion distributions of groups in two empirical networks and displays a bell-shaped distribution with an irregular symmetry, reaching its peak at the mainstream opinion. The analysis and simulation results of the network reveal small fluctuations that are irregular in nature. By visualizing the public opinion distribution in the network, as seen in Fig.\ref{fig:3}(g-l) and Fig.\ref{fig:8}, we observe that the distribution is closely related to the network's community structure, meaning that individuals in the same community tend to hold similar opinions.
Another notable observation is that the mainstream public opinion in the population depicted in Fig~\ref{fig:2} and Fig~\ref{fig:3} tends to be consistent with the opinions of extreme groups that occupy advantageous positions. That is, if the core area is dominated by more extreme supporters, the mainstream opinion tends to support their views, and vice versa.

\section{Consistency of opinion distribution and network structure}

The formation of an individual's opinion state is a distinct characteristic that emerges from the individual's dynamic evolution, separate from the natural topological attributes of the network structure. The traditional public opinion interaction model tends to result in uniform opinions among individuals, while the public opinion model with extreme groups will make all individuals show any intermediate state from extreme opposition to extreme support. In addition, the behavior of individuals updating their own opinions according to the opinions of their neighbors prevents significant deviations in opinion between adjacent nodes, as confirmed by the visualization results in Fig.~\ref{fig:3} and Fig.~\ref{fig:8}, both in synthetic and empirical networks.
In order to quantify this phenomenon, we propose a \emph{topological correlation index} (TCI) based on the assortative coefficient\cite{re50, re51}:
\begin{eqnarray}\label{eq11}
   r_{\theta} =\frac{\sum_{i,j}(a^O_{ij} - \frac{k_ik_j}{2M})\theta_i\theta_j}{\sum_{i,j}(k_i\delta_{ij} - \frac{k_ik_j}{2M})\theta_i\theta_j},
\end{eqnarray}
where $\delta_{ij}=1$ if $i=j$, otherwise $\delta_{ij}=0$. $r_{\theta}>0$ indicates that supporters are more inclined to contact supporters and opponents are more likely to contact opponents; however, $r_{\theta}<0$ indicates the opposite.
\begin{figure}
\centering
\includegraphics[width=\linewidth]{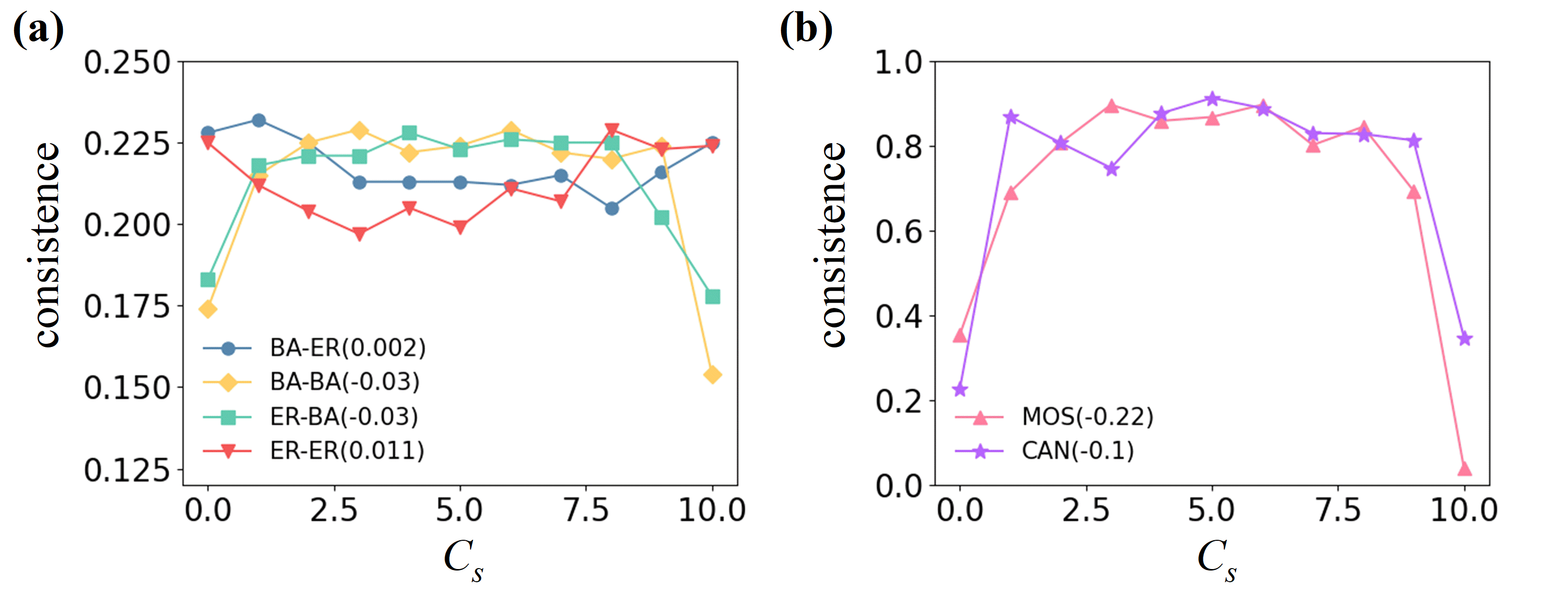}
\caption{Consistency of opinion distribution and network structure. The TCIs of our data sets under different topological distributions of the extreme group, including four synthetic networks (a): BA-ER, BA-BA, ER-BA and ER-ER, and two empirical networks (b): MOS and CAN.
In addition, the numbers in brackets represent degree assortative coefficient for each network.}\label{fig:4}
\end{figure}

Fig.~\ref{fig:4} showcases the correlation between opinion distribution and network structure in six networks with different topological distributions of extreme groups. In this experiment, we assume that $C_s$ ranges from 0 to 10. We find that the TCIs of the four synthetic networks are approximately 0.2, while the TCIs of MOS and CAN are generally above 0.6, indicating a strong alignment between the topological distribution and the opinion distribution. As for the stationary of public
opinion distribution, individuals are surrounded by people with similar opinions, which verifies the steady-state conditions for the evolution of public opinion.

\section{Relationship between mainstream opinion and node influence}

Mainstream public opinion, which represents the average of group public opinion, is often considered a crucial metric for understanding the overall tendency of group public opinion. In terms of Eq.~(\ref{eq9}), we obtain an accurate quantitative analysis of the stationary distribution of public opinion. In this section, we will discuss the relationship between the steady state of mainstream public opinion (represented by the average opinion $\bar{\theta}$) and extreme individuals.

Assume that $\rho_k^+$ and $\rho_k^-$ represent the numbers of extreme support and extreme opponent nodes with degree $k$ respectively, and $\tilde{\Gamma}_i^t$ contains individual $i$'s neighbors with state $\mathcal{U}$ in opinion layer at time $t$. Then, the evolution rule of public opinion can be written as
\begin{eqnarray}\label{eq12}
    \begin{aligned}
        \theta_{(i|d)}^{t+1} &= \frac{1}{\|\tilde{\Gamma}_i^t\|} \left( \sum_{j_g \in \Gamma_i^t}\theta_{j_g}^{t} + \sum_{j_e^+ \in \Gamma_i^t} \theta_{j_e^+}^{t} + \sum_{j_e^- \in \Gamma_i^t}\theta_{j_e^-}^{t} \right)\\
        &= \sum_{k = 1}^{N} \frac{r_k}{\|\tilde{\Gamma}_i^t\|} \sum_{j_g \in \Gamma_i^t} \left(\theta_{(j_g|k)}^{t} + \theta_{(j_e^+| k)}^{t} + \theta_{(j_e^-|k)}^{t} \right)\\
        &= \sum_{k = 1}^{N} \frac{r_k}{\|\tilde{\Gamma}_i^t\|}  \sum_{j \in \Gamma_i^t}  \left( \frac{n_k - \rho_k^+ - \rho_k^-}{n_k}\theta_{(j|k)}^{t} + \frac{\rho_k^+}{n_k} - \frac{\rho_k^-}{n_k}\right),
    \end{aligned}
\end{eqnarray}
where $n_k$ is the number of nodes with degree $k$, and $r_k = \frac{kn_k}{N \langle k \rangle}$ is the probability of a node $j$ connected to a node $i$ with degree $k$. Assume $E_d(\theta)$ is the expectation of individuals' stationary opinions whose degrees are $d$. According to the quenched mean-field method, it satisfies
\begin{eqnarray}\label{eq13}
    \begin{aligned}
        E_d(\theta) &= \sum_{k = 1}^{N} \frac{k}{N\langle k \rangle} \left[ (n_k - \rho_k^+ - \rho_k^-)E_k(\theta) \right] \\
        &\qquad + \sum_{k = 1}^{N} \frac{k}{N\langle k \rangle} \left( \rho_k^+ - \rho_k^- \right) \\
        &= \frac{1}{N} \left[ \widetilde{E}\left((n_k - \rho_k^+ - \rho_k^-)E_k(\theta) \right) + \widetilde{E}( \rho_k^+ - \rho_k^-) \right],
    \end{aligned}
\end{eqnarray}
where $\widetilde{E}(x) = \sum_{k = 1}^{N} kx / \langle k \rangle$.
From Eq.~(\ref{eq13}), we can see $E_d(\theta)$ is consistent regardless of $d$. Without loss of generality, we relabel $E_d(\theta)$ as $\bar{\theta}$ and rewrite Eq.~(\ref{eq13}) in the following way,
\begin{eqnarray}\label{eq14}
    \begin{aligned}
        \bar{\theta} &= \sum_{k = 1}^{N} \frac{k}{N\langle k \rangle} \left[ (n_k - \rho_k^+ - \rho_k^-)\bar{\theta} + \rho_k^+ - \rho_k^- \right] \\
        &= \bar{\theta} + (1 - \bar{\theta})\widetilde{E}(\rho_k^+) - (1 + \bar{\theta})\widetilde{E}(\rho_k^-).
    \end{aligned}
\end{eqnarray}
So we derive an analytical approach for the mainstream opinion:
\begin{eqnarray} \label{eq15}
\bar{\theta} = \frac{\widetilde{E}(\rho_k^+) - \widetilde{E}(\rho_k^-)}{\widetilde{E}(\rho_k^+) + \widetilde{E}(\rho_k^-)}.
\end{eqnarray}

In addition, we can derive the influence of extreme individuals on the changes of mainstream opinion. When the number of extreme supporters with degree $k$ decreases by one, mainstream opinion (denoted as $ \bar{\theta^{'}}$) satisfies,
\begin{eqnarray} \label{eq16}
\bar{\theta^{'}} = \frac{\widetilde{E}(\rho_k^+) - \widetilde{E}(\rho_k^-) - k}{\widetilde{E}(\rho_k^+) + \widetilde{E}(\rho_k^-) - k}.
\end{eqnarray}
Therefore, let $\bar{\theta}_k$ represent the difference of mainstream opinion when the number of extreme supporters with degree $k$ is reduced by one, satisfying
\begin{eqnarray} \label{eq17}
 \begin{aligned}
    \Delta \bar{\theta}_k &= \bar{\theta} - \bar{\theta^{'}}\\
    &= \frac{2 \widetilde{E}(\rho_k^-)}{\widetilde{E}(\rho_k^+) + \widetilde{E}(\rho_k^-)} \cdot \frac{k}{ \widetilde{E}(\rho_k^+) + \widetilde{E}(\rho_k^-) - k}\\
    &\sim \frac{k}{ \widetilde{E}(\rho_k^+) + \widetilde{E}(\rho_k^-) - k}.
 \end{aligned}
\end{eqnarray}

Theoretical analysis based on the information-public opinion model indicates that mainstream public opinion is not only the peak of group public opinion distribution, but also influenced by the location of extreme groups. For a fixed network topology and fixed dynamic parameters, the degree of an extreme individual is the sole factor that determines the extent to which it impacts mainstream opinion. The topological distribution of extreme groups plays a crucial role in the steady state of public opinion distribution.
As indicated by Eq.~(\ref{eq17}), the more extreme individuals occupy hub nodes, the more mainstream opinion will tend towards their extreme opinions. This means that the impact of central extreme individuals and peripheral extreme individuals on mainstream opinion is different.

To further validate these conclusions, we design two different extreme group topology distribution schemes: an \emph{experimental scheme} and a \emph{control scheme}. In both schemes, we set $C_o = L_o$ and $F_o = 0$. The control scheme sets $C_s = 0$ and $F_s \in [0, L_s]$, so $\widetilde{E}(\rho_k^+) = F_s \langle k_{\textrm{frt}}^{F_s} \rangle / \langle k \rangle$ and $\widetilde{E}(\rho_k^-) = L_o \langle k_{\textrm{ctr}}^{L_o} \rangle / \langle k \rangle$.
While in the experimental scheme, we guarantee $C_s = F_s \in [0, L_s / 2]$. Therefore,
$\widetilde{E}(\rho_k^+) = [C_s\langle k_{\textrm{ctr}}^{C_s} \rangle + F_s\langle k_{\textrm{frt}}^{F_s} \rangle] / \langle k \rangle$ and $\widetilde{E}(\rho_k^-) = C_0 \langle k_{\textrm{ctr}}^{C_0} \rangle / \langle k \rangle$.
Obviously, it satisfies $C_s \langle k_{\textrm{ctr}}^{C_s} \rangle + C_o \langle k_{\textrm{ctr}}^{C_o} \rangle = (C_s + C_o) \langle k_{\textrm{ctr}}^{C_s + C_o} \rangle$. Furthermore, we randomly select $C_s$ central extreme supporters from $C_s + C_o$ central nodes, it means the expectation of the average degree of the $C_s$ extreme supporters is equal to the average degree of the $C_s + C_o$ hub nodes, that is, $\langle k_{\textrm{ctr}}^{C_s} \rangle \approx \langle k_{\textrm{ctr}}^{C_o} \rangle \approx \langle k_{\textrm{ctr}}^{C_s + C_o} \rangle$.
According to Eq.~(\ref{eq15}), the mainstream opinion satisfies
\begin{eqnarray}\label{eq18}
\bar{\theta} = \left\{
\begin{aligned}
&\frac{F_s \langle k_{\textrm{frt}}^{F_s} \rangle - L_o \langle k_{\textrm{ctr}}^{L_o} \rangle}{F_s \langle k_{\textrm{frt}}^{F_s} \rangle + L_o \langle k_{\textrm{ctr}}^{L_o} \rangle},\qquad\qquad\quad \text{\emph{control}}\\
&\frac{F_s \langle k_{\textrm{frt}}^{F_s} \rangle + (F_s - L_o) \langle k_{\textrm{ctr}}^{F_s + L_o} \rangle}{F_s \langle k_{\textrm{frt}}^{F_s} \rangle + (F_s + L_o) \langle k_{\textrm{ctr}}^{F_s + L_o}\rangle}, \quad\text{\emph{experimental}}
\end{aligned}
\right.
\end{eqnarray}
where $\langle k_{\textrm{ctr}}^{F_s} \rangle$ and $\langle k_{\textrm{frt}}^{F_s} \rangle$ represent the average degree of the $F_s$ central extreme nodes and $F_s$ frontier extreme nodes, respectively.

\begin{figure}[h]
\centering
\includegraphics[width=\linewidth]{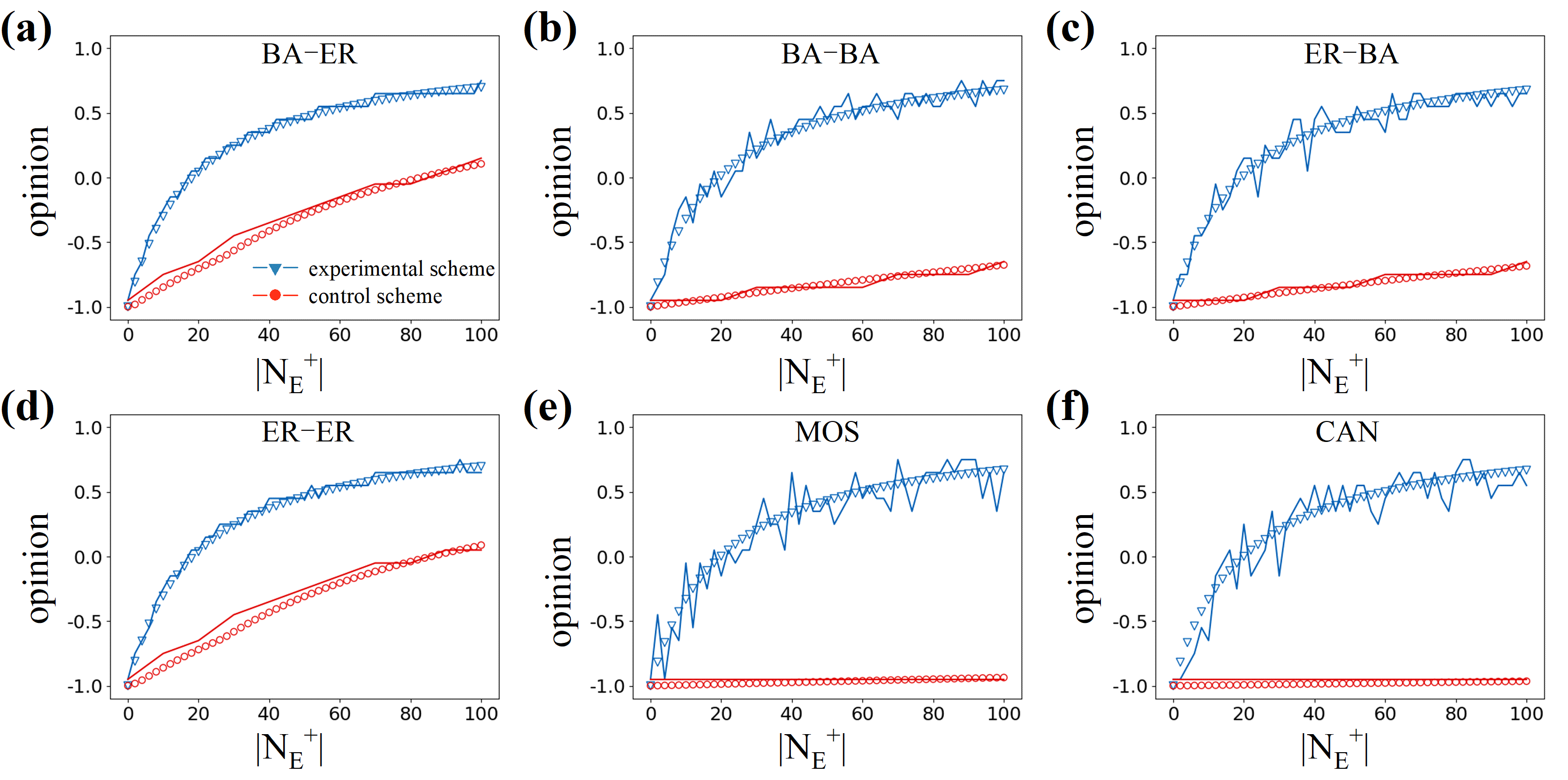}
\caption{Extreme supporters that occupy the hub nodes play a greater role in
the traction of mainstream opinion. The mainstream opinion in control scheme (red) and experimental scheme (blue) based on our data set, including BA-ER (a), BA-BA (b), ER-BA (c), ER-ER (d), MOS (e) and CAN (f).
The lines and points represent the simulation results and analytical results.}\label{fig:5}
\end{figure}

The results of both the analysis and simulation of the two schemes, as shown in Fig.~\ref{fig:5}, demonstrate the impact of extreme supporters on the mainstream opinion, where $L_o = 10$ and $L_s = 100$. The control scheme reveals that the mainstream opinion gradually moves towards the state of support as the number of extreme supporters increases, while the experimental scheme exhibits a quicker shift towards the extreme support state. Our findings, based on both synthetic and empirical networks, indicate that extreme supporters located at hub nodes have a greater influence on the mainstream opinion. This highlights the potential for designing public opinion regulation strategies based on the topological distribution of extreme groups.

\section{Public opinion regulation strategies}
The analysis using the quenched mean field method provides a fresh interpretation of how mainstream opinion is formed based on the distribution of extreme individuals.
To further clarify the relationship between extreme individuals and mainstream opinions,
for ease of calculation, we assume that the number of extreme individuals on both sides is equal, that is, $|N_E^+| = |N_E^-| = L$.
Then, according to Eq.~(\ref{eq15}), the relationship between mainstream opinion $\bar{\theta}$ and $C_s$ is as follows,
\begin{eqnarray} \label{eq19}
\bar{ \theta} = \frac{2C_s - L}{L} \cdot \frac{\langle k_{\textrm{ctr}}^L \rangle - \langle k_{\textrm{frt}}^L \rangle}{\langle k_{\textrm{ctr}}^L \rangle + \langle k_{\textrm{frt}}^L \rangle}.
\end{eqnarray}
For large-scale networks with a power-law degree distribution, Eq.~(\ref{eq19}) can be simplified
\begin{eqnarray} \label{eq20}
\bar{ \theta} \sim \frac{2C_s}{L} - 1.
\end{eqnarray}

\begin{figure}[h]
\centering
\includegraphics[width=\linewidth]{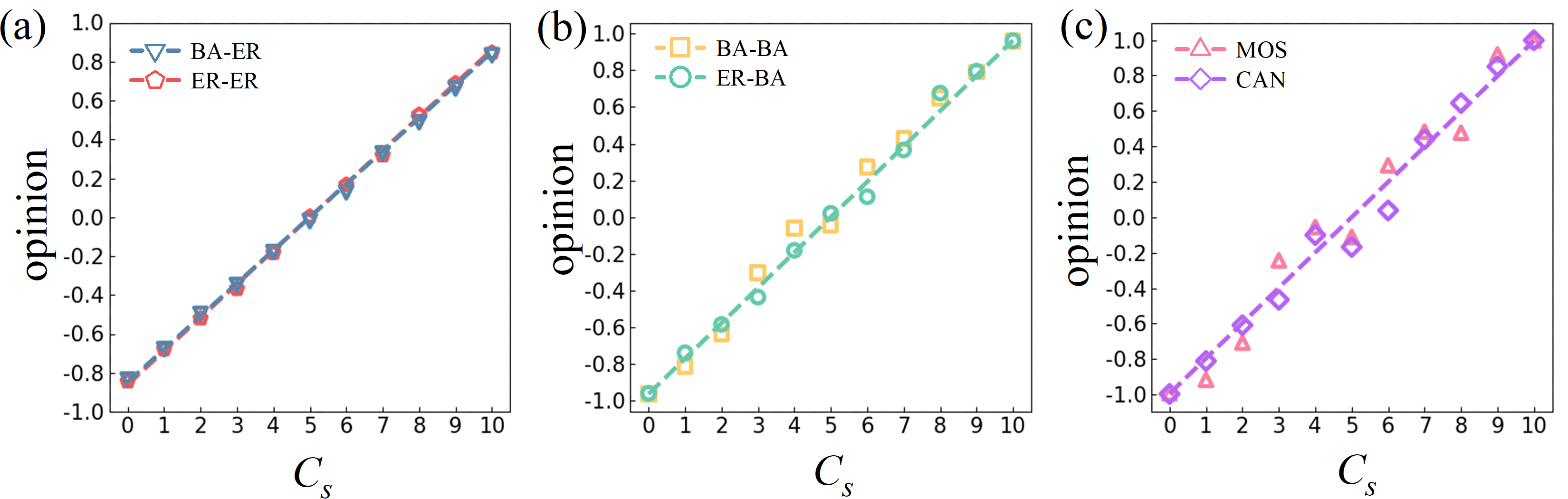}
\caption{The effectiveness of opinion regular strategy: theory versus simulation.
 Each data point represents the simulation value of mainstream opinion and each dotted line is the analytic result according to Eq.~(\ref{eq19}).
 (a): the results on two synthetic networks: BA-ER and ER-ER.
 (b): the results on two synthetic networks: BA-BA and ER-BA.
 (c): the results on two empirical networks: MOS and CAN.}
\label{fig:6}
\end{figure}

According to Eq.(\ref{eq19}) and Eq.(\ref{eq20}), a clear linear relationship is established between the mainstream opinion and the number of central extreme individuals. This relationship is further confirmed by the simulation results on both synthetic and empirical networks, as shown in Fig.~\ref{fig:6}.
This means that the mainstream opinion can be controlled and adjusted to any point between extreme opposition and support by adjusting the number of central extreme individuals.
Most importantly, this opens up the possibility to regulate the impact of extreme individuals on group public opinion, which can help mitigate the adverse effects of social media on public health policies, ensure the effective implementation of such policies, and ultimately enhance public health outcomes.

\section{Discussion}

We propose a novel information-opinion model that incorporates timely information dissemination and dynamic opinion dynamics on a duplex network, to investigate the mechanisms behind public opinion evolution and develop regulation strategies at a theoretical level. Our study considers a heterogeneous population consisting of two types of extreme individuals who maintain their extreme opinions.
We theoretically reveal the influence mechanism of these extreme individuals on public opinion, finding a bell-shaped distribution of the majority of individual opinions, which are centered around the mainstream public opinion.
Additionally, the distribution of public opinion is topologically correlated with the network, leading to the formation of small groups of supporters or opponents within the network.
Moreover, our analysis highlights the relationship between mainstream opinion and node influence, revealing a significant difference in the impact of extreme individuals in the hub area and the frontier area on the mainstream opinion.
Based on these results, we propose a public opinion regulation strategy, both theoretically and experimentally. Our findings provide valuable insights into the interaction of various dynamic processes and the regulation of public opinion and social contagion. These insights could serve as a basis for implementing health policies influenced by public opinion.

Understanding the underlying mechanisms of information dissemination and public opinion interaction is essential in modern social science, especially with the rapid advancement of media technology. Thus, regulating public opinion to promote healthy development and reduce the spread of rumors and extreme comments has become a priority.
Despite the simplicity of our model assumptions, our research provides insights into the influence mechanism of extreme individuals on public opinion. The interaction process and theoretical analysis methods we have proposed can also be extended to other areas, such as human behavior and disease transmission dynamics.
Furthermore, in the next step of our study, we aim to integrate real-world data into our model to further expand the framework of our theoretical analysis.
Our theoretical framework and public opinion regulation strategy, viewed through the lens of a multi-layered complex network, offer a new way of exploring the interplay of dynamic network mechanisms and shaping network public opinion.

\begin{acknowledgments}
We thank Fan-peng Song for useful comments on early draft. The authors are supported by National Natural Science Foundation of China under Grant 12001324, Grant 12231018, Grant 11971270.
\end{acknowledgments}

\appendix

\section{Data description}\label{apx:1}
We use the methods proposed above to analyze the public opinion distribution of several synthetic and empirical networks:
\begin{itemize}
\item Four synthetic networks: In this paper, we use four two-layer synthetic network models consisting of pairs of Barab\'{a}si-Albert network (BA) \cite{re52} and Erd\H{o}s-R\'{e}nyi network (ER) \cite{re53} models, namely BA-ER, BA-BA, ER-ER and ER-BA. Each layer contains 10,000 nodes with an average degree of 10.
\item Two empirical networks:
    We have built two empirical networks, known as MOS and CAN, based on the data gathered from real-world information dissemination and opinion interaction processes \cite{re54}. We have taken into account the information exchange between users ("mention" layer  in \cite{re54}) and labeled it as the information layer and the communication between users ( "reply" layer in \cite{re54}), labeled as the opinion layer. 

    MOS: It reflects the different social relations among users during the 2013 World Championships in Athletics, including 88804 users and 210250 connections. In this network, each user represents a node, and a connection $(i, j)$ represents an edge connecting user $i$ and $j$.

    CAN: It reflects the different social relations among users during the Cannes Film Festival in 2013, including 438537 users and 991854 connections. In this network, each user represents a node, and a connection $(i, j)$ represents an edge connecting user $i$ and $j$.
\end{itemize}


\section{Supplementary tables and figures}\label{apx:2}
\begin{figure}[h]
\centering
\includegraphics[width=\linewidth]{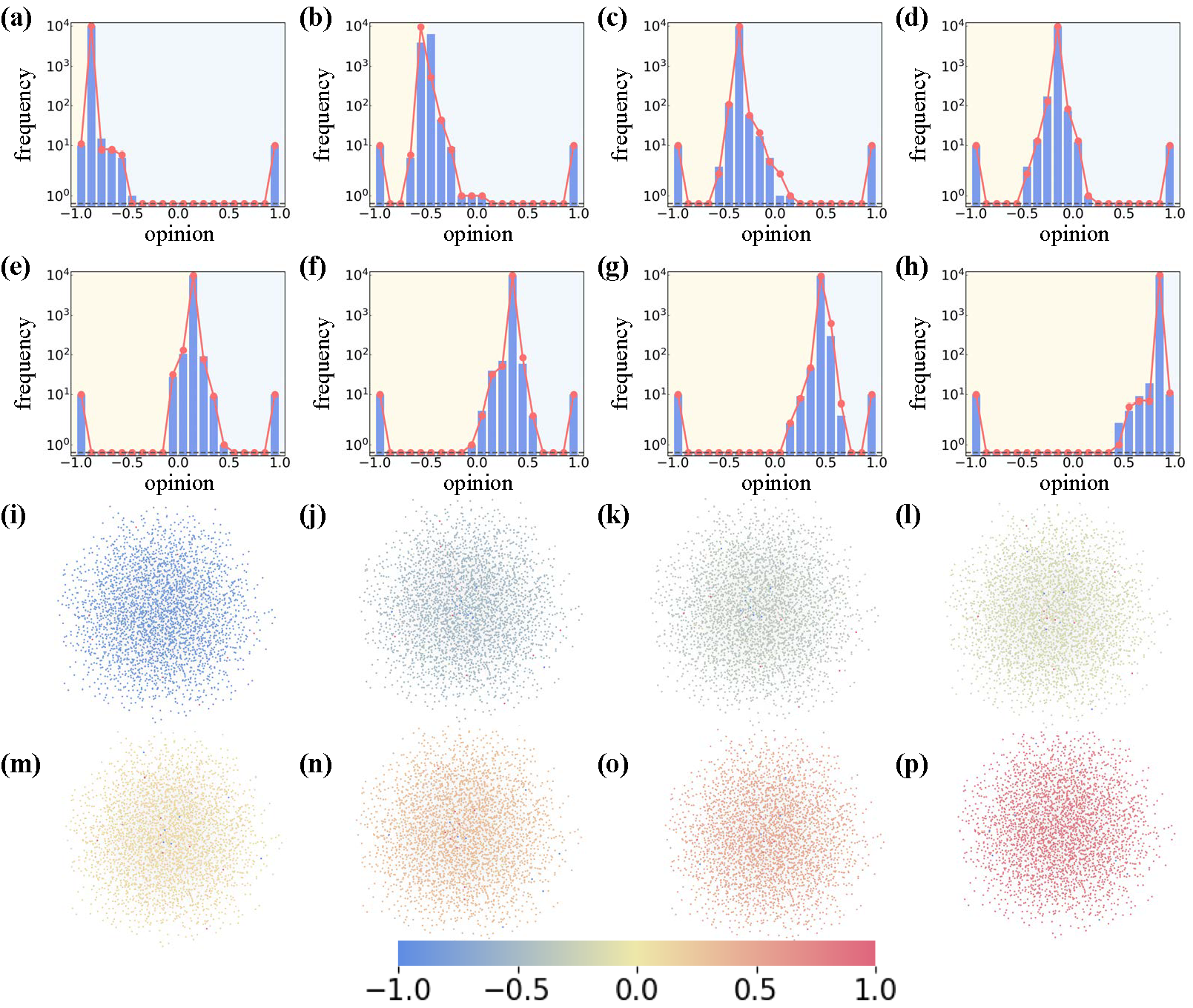}
\caption{Frequency statistics and visualization of public opinion on BA-ER. The parameters~$\omega$,~$\gamma$~and~$h$~are set as ~$10$,~$0.05$~and~$1$, respectively. This figure shows the results based on eight cases: $C_s$ = 0 (a,i), $C_s$ = 2 (b,j), $C_s$ = 3 (c,k), $C_s$ = 4 (d,l), $C_s$ = 6 (e,m), $C_s$ = 7 (f,n), $C_s$ = 8 (g,o) and $C_s$ = 10 (h,p).}
\label{fig:9}
\end{figure}

\begin{figure}
\centering
\includegraphics[width=\linewidth]{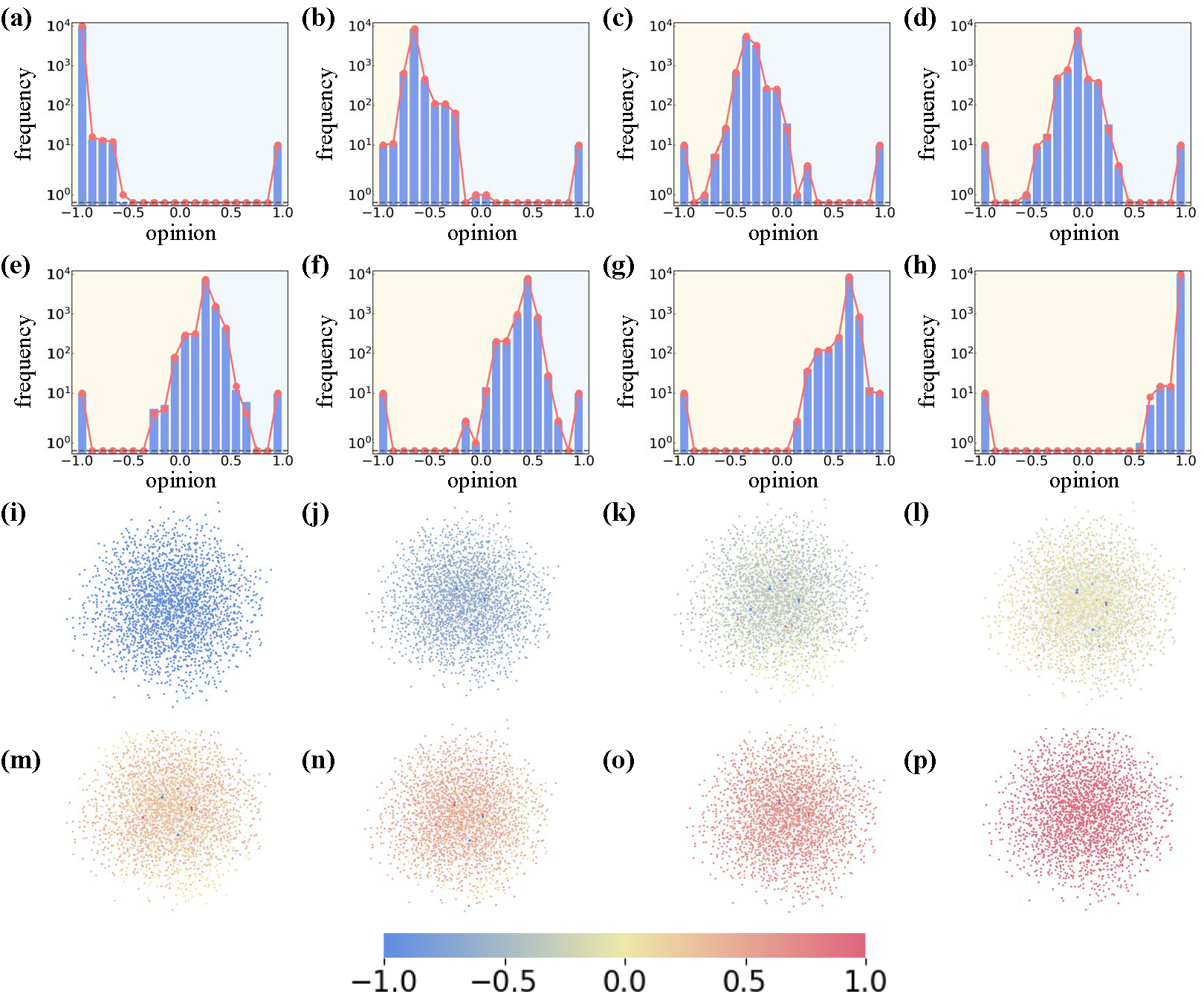}
\caption{Frequency statistics and visualization of public opinion on BA-BA. The parameters~$\omega$,~$\gamma$~and~$h$~are set as ~$10$,~$0.05$~and~$1$, respectively. This figure shows the results based on eight cases: $C_s$ = 0 (a,i), $C_s$ = 2 (b,j), $C_s$ = 3 (c,k), $C_s$ = 4 (d,l), $C_s$ = 6 (e,m), $C_s$ = 7 (f,n), $C_s$ = 8 (g,o) and $C_s$ = 10 (h,p).}
\label{fig:10}
\end{figure}

\begin{figure}
\centering
\includegraphics[width=\linewidth]{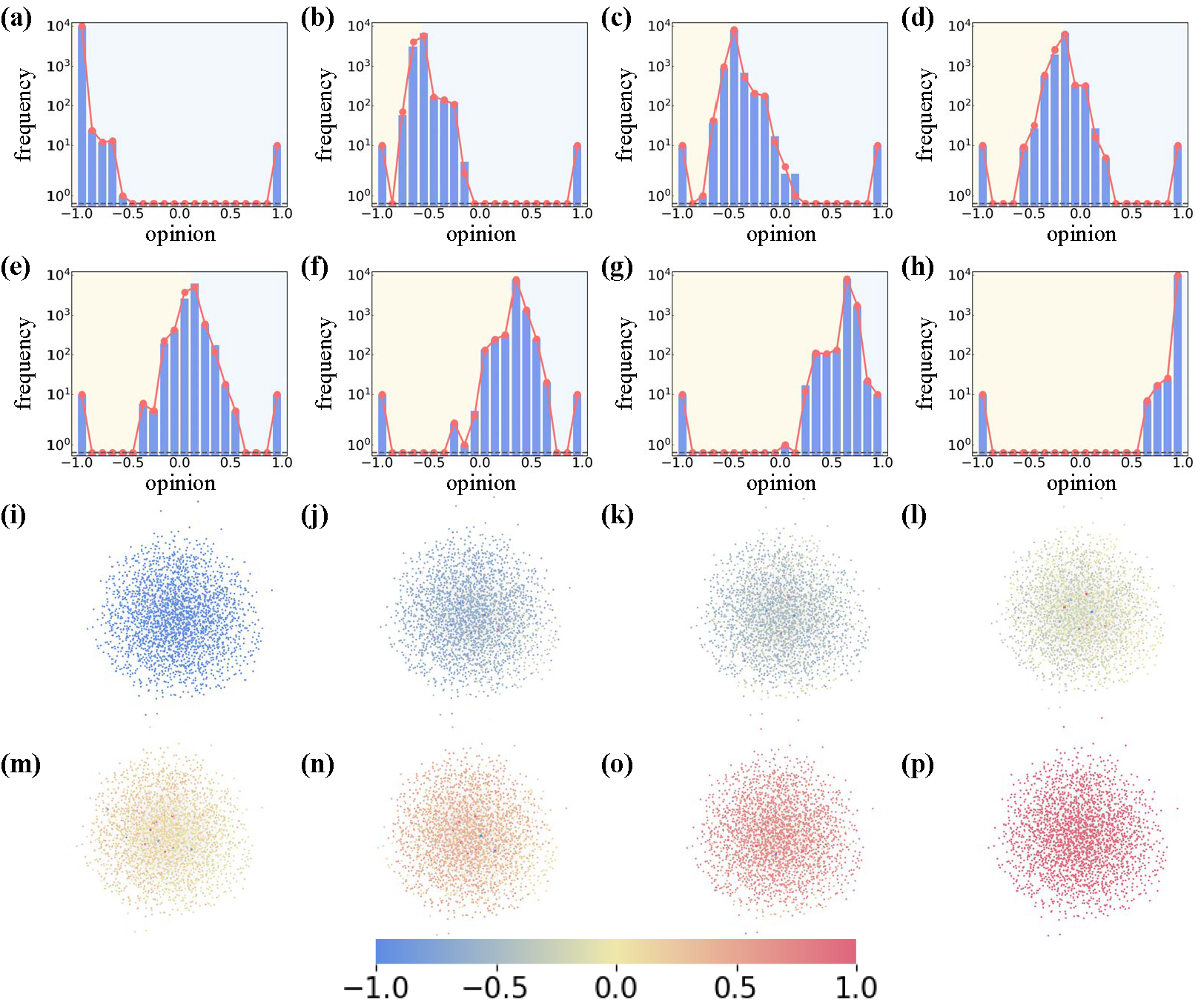}
\caption{Frequency statistics and visualization of public opinion on ER-BA. The parameters~$\omega$,~$\gamma$~and~$h$~are set as ~$10$,~$0.05$~and~$1$, respectively. This figure shows the results based on eight cases: $C_s$ = 0 (a,i), $C_s$ = 2 (b,j), $C_s$ = 3 (c,k), $C_s$ = 4 (d,l), $C_s$ = 6 (e,m), $C_s$ = 7 (f,n), $C_s$ = 8 (g,o) and $C_s$ = 10 (h,p).}
\label{fig:11}
\end{figure}

\begin{figure}
\centering
\includegraphics[width=\linewidth]{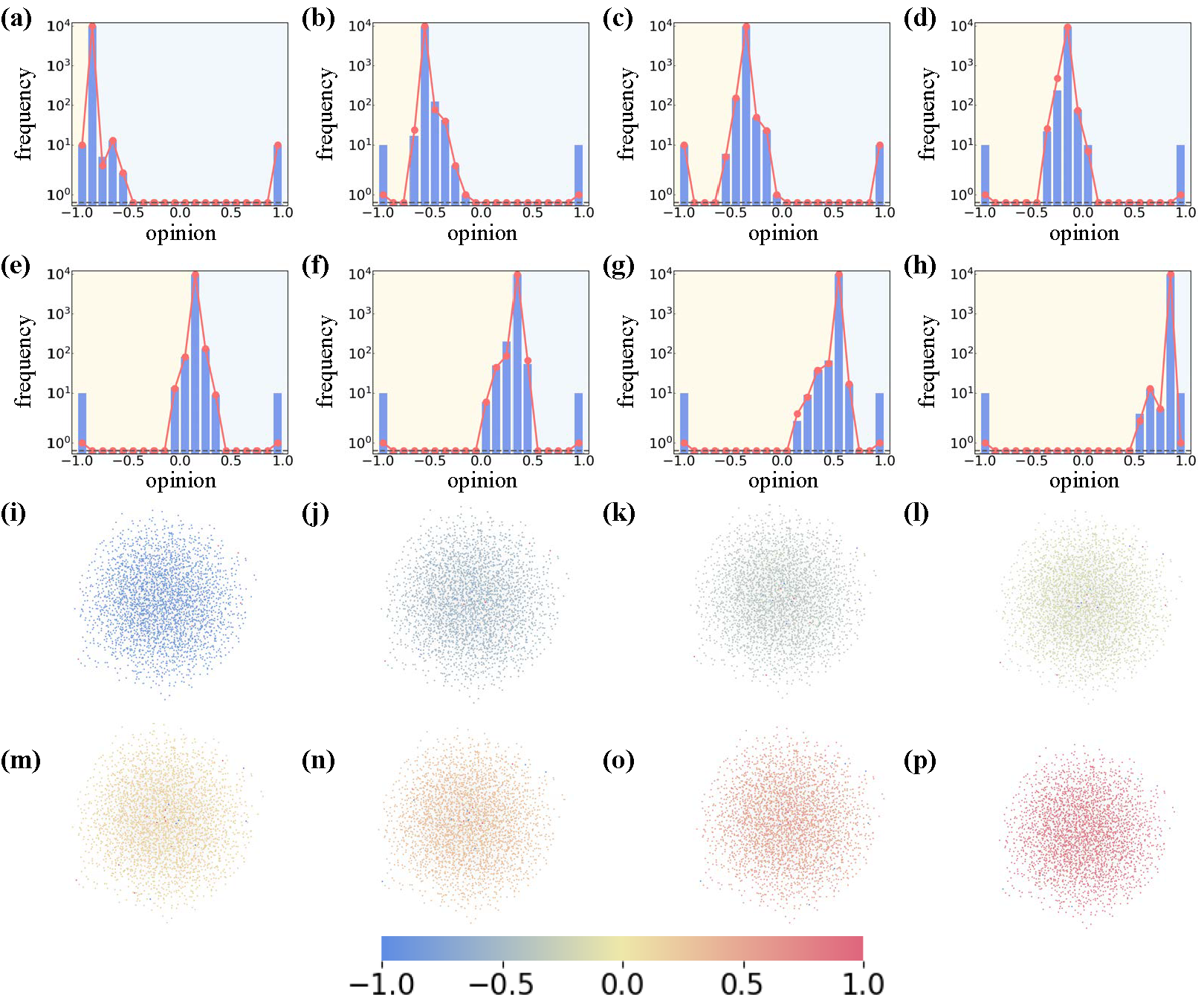}
\caption{Frequency statistics and visualization of public opinion on ER-ER. The parameters~$\omega$,~$\gamma$~and~$h$~are set as ~$10$,~$0.05$~and~$1$, respectively. This figure shows the results based on eight cases: $C_s$ = 0 (a,i), $C_s$ = 2 (b,j), $C_s$ = 3 (c,k), $C_s$ = 4 (d,l), $C_s$ = 6 (e,m), $C_s$ = 7 (f,n), $C_s$ = 8 (g,o) and $C_s$ = 10 (h,p).}
\label{fig:12}
\end{figure}

\begin{figure}
\centering
\includegraphics[width=\linewidth]{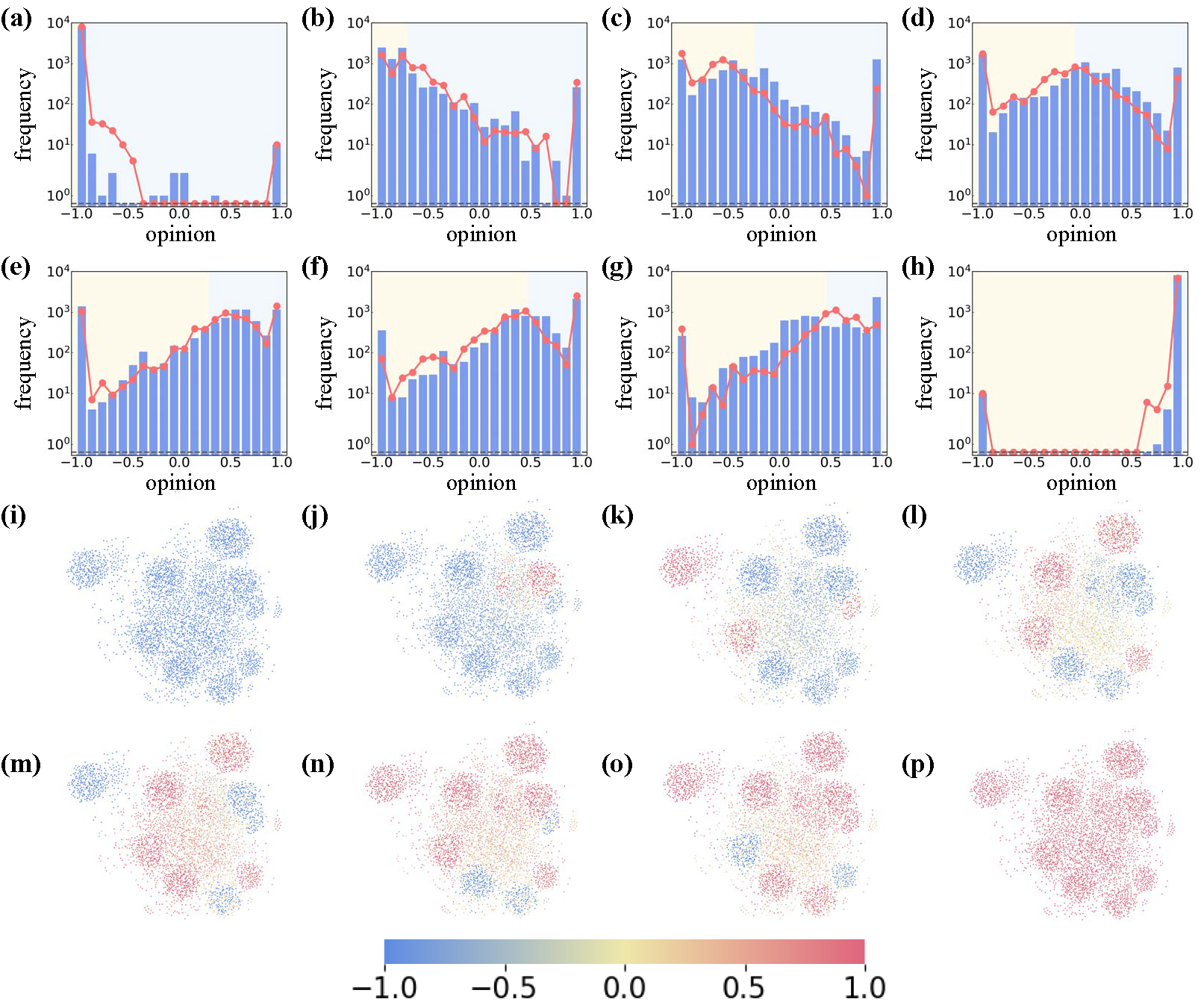}
\caption{Frequency statistics and visualization of public opinion on MOS. The parameters~$\omega$,~$\gamma$~and~$h$~are set as ~$10$,~$0.05$~and~$1$, respectively. This figure shows the results based on eight cases: $C_s$ = 0 (a,i), $C_s$ = 2 (b,j), $C_s$ = 3 (c,k), $C_s$ = 4 (d,l), $C_s$ = 6 (e,m), $C_s$ = 7 (f,n), $C_s$ = 8 (g,o) and $C_s$ = 10 (h,p).}
\label{fig:13}
\end{figure}

\begin{figure}
\centering
\includegraphics[width=\linewidth]{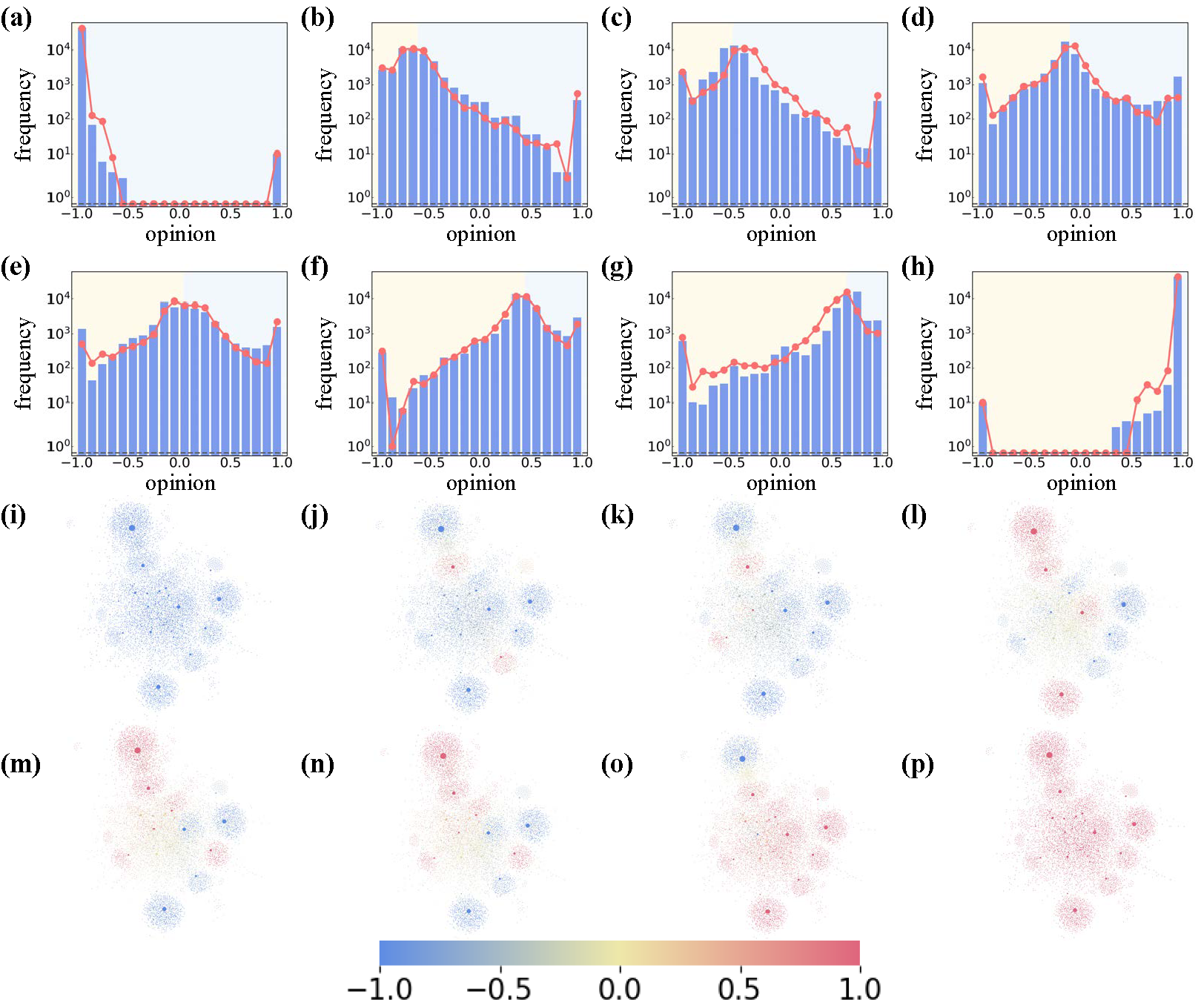}
\caption{Frequency statistics and visualization of public opinion on CAN. The parameters~$\omega$,~$\gamma$~and~$h$~are set as ~$10$,~$0.05$~and~$1$, respectively. This figure shows the results based on eight cases: $C_s$ = 0 (a,i), $C_s$ = 2 (b,j), $C_s$ = 3 (c,k), $C_s$ = 4 (d,l), $C_s$ = 6 (e,m), $C_s$ = 7 (f,n), $C_s$ = 8 (g,o) and $C_s$ = 10 (h,p).}
\label{fig:14}
\end{figure}

\begin{figure}
\centering
\includegraphics[width=\linewidth]{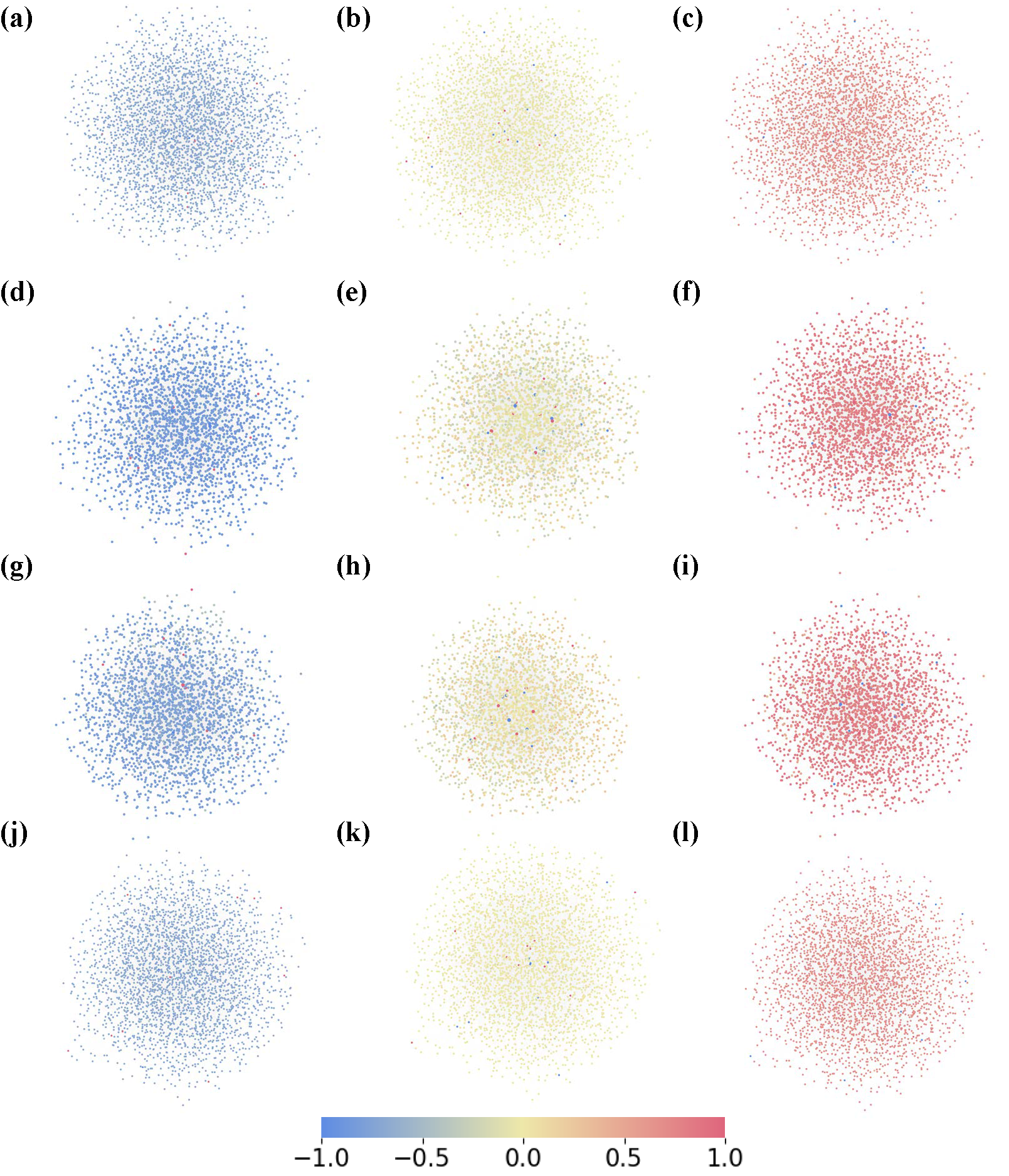}
\caption{Visualization of public opinion on four synthetic networks in Fig~\ref{fig:2}, including BA-ER (a-c), BA-BA (d-f), ER-BA (g-i) and ER-ER (j-l). The parameters~$\omega$,~$\gamma$~and~$h$~are set as ~$10$,~$0.05$~and~$1$, respectively. This figure shows the results based on three cases: $C_s=1$ (a, d, g, j), $C_s=5$ (b, e, h, l) and $C_s=9$ (c, f, i, l).}
\label{fig:8}
\end{figure}

The main notations in paper is shown in Table~\ref{Tab1}.
More, we show other simulation and theoretical results for BA-ER (Fig.~\ref{fig:9}), BA-BA (Fig.~\ref{fig:10}), ER-BA (Fig.~\ref{fig:11}), ER-ER (Fig.~\ref{fig:12}), MOS (Fig.~\ref{fig:13}) and CAN (Fig.~\ref{fig:14}), and the visualization of public opinion on four synthetic networks (Fig.~\ref{fig:8}).
\begin{table}[h]
\centering
\setlength{\abovecaptionskip}{0pt}%
\setlength{\belowcaptionskip}{10pt}%
\caption{Definitions of notations in paper}\label{Tab1}
\begin{ruledtabular}
\begin{tabular}{cc}
\centering
Notation & Description\\
\colrule
\\
$\beta_i^t$ & The spread probability of individual $i$ at time $t$\\
 & \\
$\alpha_i^t$ & The information state of individual $i$ at time $t$\\
 & \\
$\theta_i^t$ & The opinion state of individual $i$ at time $t$\\
 & \\
$\bar{\theta}$ & Mainstream opinion\\
 & \\
$C_{x}$& $x = \{s, o\}$, the numbers of central\\
 &  extreme supporters($s$) or opponents($o$)\\
 & \\
$F_{x}$& $x = \{s, o\}$, the numbers of frontier\\
 &  extreme supporters($s$) or opponents($o$)\\
 & \\
$\langle k_{\textrm{id}}^y \rangle $& $id = \{ctr, frt\}$, The average degree of the $y$\\
 &  central ($ctr$) or frontier ($frt$) extreme nodes\\
& \\
$L$& The size of extreme group\\
\end{tabular}\label{tab2}
\end{ruledtabular}
\end{table}

\clearpage
\bibliography{ref_sample}

\end{document}